\documentclass[aps,prb,twocolumn,showpacs,superscriptaddress]{revtex4-1}
\usepackage{graphicx}
\usepackage{bm}
\usepackage{color,soul}

\begin{document}

\title{Effect of Zn doping on the antiferromagnetism in kagome Cu$_{4-x}$Zn$_x$(OH)$_6$FBr}
\author{Zili Feng}
\affiliation{Beijing National Laboratory for Condensed Matter Physics, Institute of Physics, Chinese Academy of Sciences, Beijing 100190, China}
\affiliation{School of Physical Sciences, University of Chinese Academy of Sciences, Beijing 100190, China}
\author{Yuan Wei}
\affiliation{Beijing National Laboratory for Condensed Matter Physics, Institute of Physics, Chinese Academy of Sciences, Beijing 100190, China}
\affiliation{School of Physical Sciences, University of Chinese Academy of Sciences, Beijing 100190, China}
\author{Ran Liu}
\affiliation{Beijing National Laboratory for Condensed Matter Physics, Institute of Physics, Chinese Academy of Sciences, Beijing 100190, China}
\author{Dayu Yan}
\affiliation{Beijing National Laboratory for Condensed Matter Physics, Institute of Physics, Chinese Academy of Sciences, Beijing 100190, China}
\affiliation{School of Physical Sciences, University of Chinese Academy of Sciences, Beijing 100190, China}
\author{Yan-Cheng Wang}
\affiliation{Beijing National Laboratory for Condensed Matter Physics, Institute of Physics, Chinese Academy of Sciences, Beijing 100190, China}
\affiliation{School of Physical Science and Technology, China University of Mining and Technology, Xuzhou 221116, China}
\author{Jianlin Luo}
\affiliation{Beijing National Laboratory for Condensed Matter Physics, Institute of Physics, Chinese Academy of Sciences, Beijing 100190, China}
\affiliation{School of Physical Sciences, University of Chinese Academy of Sciences, Beijing 100190, China}
\affiliation{Collaborative Innovation Center of Quantum Matter, Beijing 100190, China}
\author{Anatoliy Senyshyn}
\affiliation{Heinz Maier-Leibnitz Zentrum (MLZ), Technische Universit$\ddot{a}$t M$\ddot{u}$nchen, Garching D-85747, Germany}
\author{Clarina dela Cruz}
\affiliation{Neutron Scattering Division, Neutron Sciences Directorate, Oak Ridge National Laboratory, Oak Ridge, Tennessee 37831, USA}
\author{Wei Yi}
\affiliation{Semiconductor Device Materials Group, National Institute for Materials Science, 1-1 Namiki, Tsukuba, Ibaraki 305-0044, Japan}
\author{Jia-Wei Mei}
\affiliation{Institute for Quantum Science and Engineering, and Department of Physics, Southern University of Science and Technology, Shenzhen 518055, China}
\author{Zi Yang Meng}
\affiliation{Beijing National Laboratory for Condensed Matter Physics, Institute of Physics, Chinese Academy of Sciences, Beijing 100190, China}
\affiliation{CAS Center of Excellence in Topological Quantum Computation, University of Chinese Academy of Sciences, Beijing 100190, China}
\author{Youguo Shi}
\email{ygshi@iphy.ac.cn}
\affiliation{Beijing National Laboratory for Condensed Matter Physics, Institute of Physics, Chinese Academy of Sciences, Beijing 100190, China}
\author{Shiliang Li}
\email{slli@iphy.ac.cn}
\affiliation{Beijing National Laboratory for Condensed Matter Physics, Institute of Physics, Chinese Academy of Sciences, Beijing 100190, China}
\affiliation{School of Physical Sciences, University of Chinese Academy of Sciences, Beijing 100190, China}
\affiliation{Collaborative Innovation Center of Quantum Matter, Beijing 100190, China}

\begin{abstract}
Barlowite Cu$_4$(OH)$_6$FBr shows three-dimensional (3D) long-range antiferromagnetism, which is fully suppressed in Cu$_3$Zn(OH)$_6$FBr with a kagome quantum spin liquid ground state. Here we report systematic studies on the evolution of magnetism in the Cu$_{4-x}$Zn$_x$(OH)$_{6}$FBr system as a function of $x$ to bridge the two limits of Cu$_4$(OH)$_6$FBr ($x$=0) and Cu$_3$Zn(OH)$_6$FBr ($x$=1). Neutron-diffraction measurements reveal a hexagonal-to-orthorhombic structural change with decreasing temperature in the $x$ = 0 sample. While confirming the 3D antiferromagnetic nature of low-temperature magnetism, the magnetic moments on some Cu$^{2+}$ sites on the kagome planes are found to be vanishingly small, suggesting strong frustration already exists in barlowite. Substitution of interlayer Cu$^{2+}$ with Zn$^{2+}$ with gradually increasing $x$ completely suppresses the bulk magnetic order at around $x$ = 0.4, but leaves a local secondary magnetic order up to $x\sim 0.8$ with a slight decrease in its transition temperature. The high-temperature magnetic susceptibility and specific heat measurements further suggest that the intrinsic magnetic properties of kagome spin liquid planes may already appear from $x>0.3$ samples. Our results reveal that the Cu$_{4-x}$Zn$_x$(OH)$_6$FBr may be the long-thought experimental playground for the systematic investigations of the quantum phase transition from a long-range antiferromagnet to a topologically ordered quantum spin liquid.
	
\end{abstract}



\maketitle

\section{introduction}

A quantum spin liquid (QSL) can be briefly described as a symmetric state without magnetic order emerging from strong quantum fluctuations in frustrated magnetic systems~\cite{BalentsL10,NormanMR16,SavaryL17}. The quantum fluctuations are usually enhanced by geometrical frustrations of magnetic ions, which are commonly seen in, e.g., triangle, kagome or pyrochlore lattices. Two-dimensional magnetic kagome lattices have attracted a lot of interests in the search for QSLs~\cite{PLee2008}. Theoretically, it has been shown that the kagome system may exhibit various ordered state and different QSL ground states \cite{SachdevS92,JiangHC08,YanS11,JiangHC12,MessioL12,PunkM14,BieriS15,IqbalY15,KumarK15,GongSS16,LiaoHJ17,MeiJW17,YCWang2017a,
YCWang2017b}, such as chiral and Z$_2$ QSL. These kagome QSLs are usually very close in energy~\cite{MendelsP16} and depend sensitively on the particular form of the superexchange couplings, which render them difficult to be tested experimentally. 

Experimental progress in finding kagome QSLs has been substantial. Among many kagome magnets, herbertsmithite ZnCu$_3$(OH)$_6$Cl$_2$ shows several promising properties of a QSL~\cite{NormanMR16}. First of all, it consists of perfect kagome Cu$^{2+}$ ($s=1/2$) planes that show no magnetic order down to at least 20 mK \cite{ShoresMP05,BertF07,MendelsP07,HeltonJS07}. Inelastic neutron scattering (INS) experiments display broad dispersionless magnetic excitations that are consistent with spinon continuum expected in QSLs \cite{HanTH12}. Later nuclear magnetic resonance (NMR) and INS experiments suggest that the system may be gapped \cite{FuM15,HanTH16}. Interestingly, previous studies have suggested that herbertsmithite may be close to a quantum critical point (QCP) \cite{HeltonJS10}. However, it is later found that the low-energy spin excitations ($<$ 1 meV) are dominated by the so-called "impurities" of residual interlayer Cu$^{2+}$ ions due to imperfect substitution of inter-kagome Cu by Zn \cite{NilsenGJ13,HanTH16}. Moreover, Cu$_4$(OH)$_6$Cl$_2$, the base material that leads to herbertsmithite, has at least four polymorphs with different nuclear structures that are all different from herbertsmithite and have different magnetic orders \cite{HawthorneFC85,PariseJB86,GriceJD96,JamborJL96,MalcherekT09,ZhengXG05,ZhengXG05b,ZhengXG05c,LeeSH07,KimJH08,WillsAS08}. INS experiments also do not support the presence of a QCP in the Zn$_x$Cu$_{4-x}$(OD)$_6$Cl$_2$ system since the antiferromagnetic (AF) order in the x = 0 sample becomes spin-glass-like with increasing x before the QSL is established in the x = 1 sample \cite{LeeSH07}.

Recently, a new compound of Cu$_3$Zn(OH)$_6$FBr has been synthesized to exhibit properties that are consistent with a Z$_2$ QSL \cite{FengZL17,WenXG17,WeiY17}. This compound is obtained by substituting interlayer Cu$^{2+}$ in barlowite Cu$_4$(OH)$_6$FBr with nonmagnetic Zn$^{2+}$. The barlowite has perfect Cu$^{2+}$ kagome planes with an AF transition at about 15 K \cite{HanTH14,JeschkeHO15,HanTH16b}. Since the barlowite and Cu$_3$Zn(OH)$_6$FBr have the same space group for the crystal structures at room temperature, it may provide us a rare opportunity to study the quantum phase transition from an AF ordered state to a QSL ground state by continuously tuning Zn substitution level. Therefore, the Cu$_{4-x}$Zn$_x$(OH)$_6$FBr system may provide the long-thought experimental playground for the investigation of novel quantum phase transitions between symmetry-breaking phases and symmetric topologically ordered phases.  

In this paper, we systematically investigate the magnetic properties of the Cu$_{4-x}$Zn$_x$(OH)$_6$FBr system. Our results suggest that Zn can indeed be continuously doped into barlowite and suppress the long-range AF order. However, a hexagonal-to-orthorhombic structural change is observed in barlowite, and the low-temperature magnetic structure is directly associated with the orthorhombic structure. According to x-ray diffraction and magnetic susceptibility measurements, we find that the interlayer Cu$^{2+}$ ions result in lattice distortion and a local magnetic order up to $x$ = 0.82. However, the specific-heat measurements suggest that the bulk three-dimensional (3D) antiferromagnetic order should disappear around $x=0.4$ and the spin dynamics of QSL kagome planes may already start to evolve from $x >0.3$, which finally leads to a gapped QSL in Cu$_3$Zn(OH)$_6$FBr. Putting these information together, our comprehensive results suggest that the Cu$_{4-x}$Zn$_x$(OH)$_6$FBr system host rich physics of the interplay of frustration, antiferromagnetic order as well as the topologically ordered QSL.

\section{experiments}

Powders of Cu$_{4-x}$Zn$_x$(OH)$_6$FBr were synthesized by a hydrothermal method. We sealed powders of Cu$_2$(OH)$_2$CO$_3$ (1.5 mmol), NH$_4$F (1 mmol), ZnBr$_2$ ($x_{nom}$ mmol) and CuBr$_2$ (1-$x_{nom}$ mmol) in a 50-ml reaction vessel with half water, where $x_{nom}$ is the nominal Zn content. The vessel was slowly heated to 200 $^{\circ}$C and kept for 12 hours before cooling down to room temperature. Powders of Cu$_{4-x}$Zn$_x$(OH)$_6$FBr were obtained by drying the products. The Zn content is determined by the inductively coupled plasma mass spectrometer (Thermo IRIS Intrepid II) with an uncertainty of about 4\%. It should be noted for the Cu$_3$Zn(OH)$_6$FBr, the Zn content is determined to be about 0.92. We will use the actual Zn substitution level throughout the paper, so the $x$ = 0.92 sample is equivalent to Cu$_3$Zn(OH)$_6$FBr reported previously \cite{FengZL17,WeiY17}. To obtain the deuterated samples, the mixture is changed to CuO, ZnF$_2$, ZnBr$_2$ and CuBr$_2$, and we use heavy water instead. The content of $H$ is less than 2\% according to the NMR measurement. The magnetic susceptibility and heat capacity were measured by the magnetic property measurement system (MPMS) and physical property measurement system (PPMS, Quantum Design), respectively. The structures of the Cu$_{4-x}$Zn$_x$(OH)$_6$FBr system were measured at room temperature by the x-ray diffractometer (Rigaku Ultima IV) with Cu K$_{\alpha}$ radiation and a scintillation counter detector. The magnetic and nuclear structures of Cu$_4$(OD)$_6$FBr are determined by neutron-diffraction experiments performed on the SPODI diffractometer at FRM-II, Germany and the HB-2A diffractometer at HFIR, USA with wavelengths of 1.5483 \AA and 2.4103 \AA, respectively. 

\section{results}

\subsection{Nuclear and magnetic structures in Cu$_4$(OH)$_6$FBr}

\begin{figure}[tbp]
\includegraphics[scale=1]{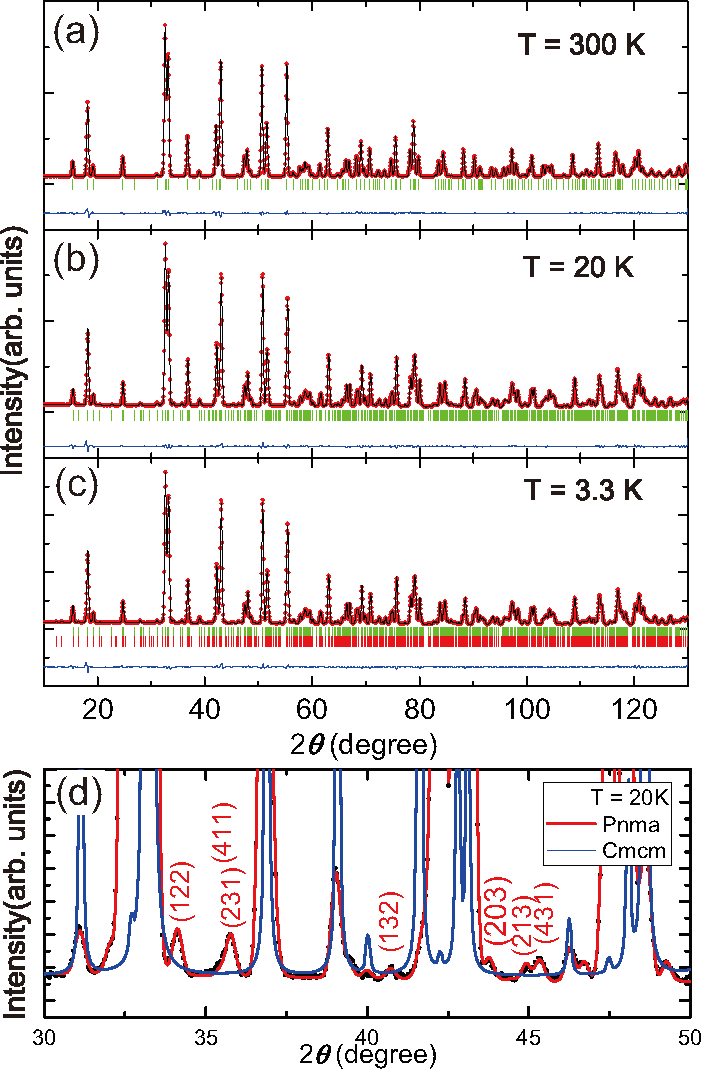}
\caption{Neutron powder diffraction intensities of Cu$_4$(OD)$_6$FBr (red dots) at (a) 300 K, (b) 20 K, and (c) 3.3 K measured at SPODI. The calculated intensities are shown by the black lines. Short vertical green and red lines represents nuclear and magnetic Bragg peak positions, respectively. The blue line shows the difference between measured and calculated intensities. The weighted profile R-factor ($R_{wp}$) is 4.22\%, 4.46\%, and 4.29\% for (a) - (c), respectively. (d) Comparison between the calculated intensities of the Pnma (red) and Cmcm (blue) structures at 20 K. The labeled peaks are in the Pnma notation and cannot be indexed in the Cmcm structure. }
\end{figure}

\begin{figure}[tbp]
\includegraphics[scale=0.75]{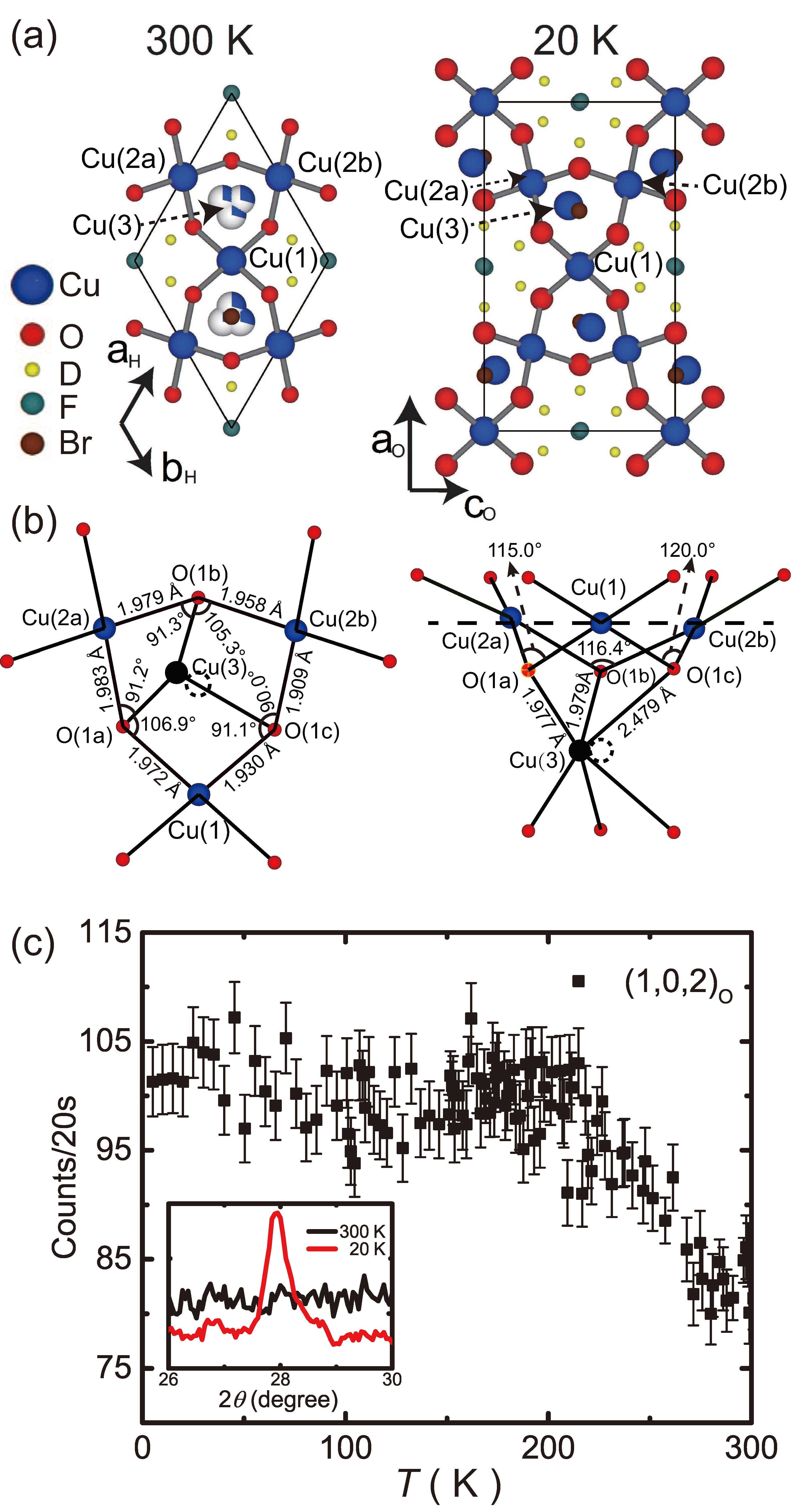}
\caption{(a) Nuclear structure at 300 K (left) and 20 K (right) in the view vertical to the kagome plane. The diamond and rectangle are in-plane unit cells at 300 and 20 K, respectively. The subscripts of "H" and "O" denote hexagonal and orthorhombic structures, respectively. We note that the c axis at room temperature becomes the b axis at low temperatures. The three overlapping Cu atoms at 300 K represent three positions that the atom may actually occupy. (b) Detailed low-temperature structure showing only Cu (large blue circles) and O (small red circles) in the view parallel to orthorhombic b (left) and orthorhombic a (right), respectively. (c) Temperature dependence of the (1,0,2)$_O$ structural peak measured at HB-2A. The inset shows 2$\theta$ scans around this peak at 20 and 300 K. }
\end{figure}

Figures 1(a)-1(c) show the neutron-diffraction and refinement results of Cu$_4$(OD)$_6$FBr at 300, 20 and 3.3 K, respectively \cite{supp1}. The refinement of the data at 3.3 K has to include an AF structure as discussed later in this section. The nuclear structure at room temperature is found to be the same as reported previously \cite{HanTH14} with parameters shown in Table I(a). At 20 K that is above $T_N$, the data can be well described by an orthorhombic structure (Pnma) as shown in Table I(b). Recently, an orthorhombic structure with a different space group Cmcm was reported for the single-crystal barlowite \cite{PascoCM18}. Figure 1(d) shows the comparison between the calculated results of these two structures at 20 K on our sample, where the Cmcm structure cannot describe the data. This discrepancy may come from the different methods used to prepare the samples.

Figure 2(a) depicts the change in the structure in the view vertical to the kagome planes. According to the low-temperature structure, we label the three Cu$^{2+}$ ions in the kagome planes as Cu(1), Cu(2a), and Cu(2b), the interlayer Cu$^{2+}$ as Cu(3). At room temperature, the first three of them are symmetrically equivalent [Cu1 in Table I(a)] and forms an equilateral triangle. Cu(3) [Cu2 in Table I(a)] has three equivalent positions with an average position at the center of the triangle, as shown in the left panel of Fig. 2(a). The distribution of Cu(3) ions is presumably random. At 20 K, this random distribution is replaced by a regular pattern as shown in the right panel of Fig. 2(a) and the position of Cu3 in Table I(b). In the meantime, the positions of Cu(2a) and Cu(2b), which are symmetrically equivalent in the Pnma space group and labeled as Cu2 in Table I(b), are distorted both within and out of the kagome plane whereas that of Cu(1) [Cu1 in Table I(b)] remains unchanged. The new unit cell is defined according to the positions of the Cu(1) ions. 

\begin{table}
  \centering
    \begin{tabular}{ccccccc}
        \hline
        (a) & Site & x & y & z & B (\AA$^2$)\\
        \hline
        Cu1 & 6g & 0.00000 & 0.50000 & 0.00000 & 1.403(2) \\
        Cu2 & 6h & 0.36993 & 0.63007(15) & 0.750000 & 1.089(5) \\
        Br & 2c & 0.66667 & 0.33333 & 0.75000 & 1.788(3) \\
        F & 2b & 0.00000 & 0.00000 & 0.750000 & 2.137(4) \\
        O & 12k & 0.20234 & 0.79766(7) & 0.90850(7) & 1.271(2) \\
        D & 12k & 0.13470 & 0.87530(7) & 0.86677(8) & 2.235(2) \\
        \hline
        \\
        \hline
        (b) & Site & x & y & z & B (\AA$^2$)\\
        \hline
        Cu1 & 4a & 0.00000 & 0.00000 & 0.00000 & 0.784(3) \\
        Cu2 & 8d & 0.25076(21) & 0.51109(17) & 0.24514(28) & 0.728(2) \\
        Cu3 & 4c & 0.18593(19) & 0.25000 & 0.05766(24) & 0.856(3) \\
        Br & 4c & 0.33072(23) & 0.25000 & 0.50431(50) & 0.766(2) \\
        F & 4c & 0.49753(30) & 0.25000 & 0.00719(70) & 1.107(3) \\
        O1 & 8d & 0.29741(16) & 0.09595(23) & 0.00115(45) & 0.753(3) \\
        O2 & 8d & 0.10228(24) & 0.09210(28) & 0.19800(41) & 0.772(4) \\
        O3 & 8d & 0.40066(25) & 0.58780(26) & 0.30230(39) & 0.755(4) \\
        D1 & 8d & 0.37649(15) & 0.13516(26) & 0.00117(39) & 1.543(4) \\
        D2 & 8d & 0.06257(25) & 0.13019(26) & 0.31516(39) & 1.619(5) \\
        \hline
    \end{tabular}
  \caption{Nuclear structure parameters of Cu$_{4}$(OD)$_{6}$FBr. (a) At 300 K, $P6_{3}/mmc$ (No. 194): $a = b = 6.6760(7)$ \AA, $c = 9.2952(13)$\AA, $\alpha = \beta = 90 ^{\circ}$, $\gamma = 120 ^{\circ}$. $Rp: 4.25\%$, $Rwp: 4.22\%$, $\chi^{2}: 7.55$. (b) At 20 K, $Pnma$ (No. 62): $a = 11.5129(42)$ \AA, $b = 9.2703(33)$ \AA, $c = 6.6801(28)$ \AA, $\alpha = \beta = \gamma = 90 ^{\circ}$ $Rp: 3.50\%$, $Rwp: 4.40\%$, $\chi^{2}: 13.2$.}
\end{table}

\begin{figure}[tbp]
\includegraphics[width=\columnwidth]{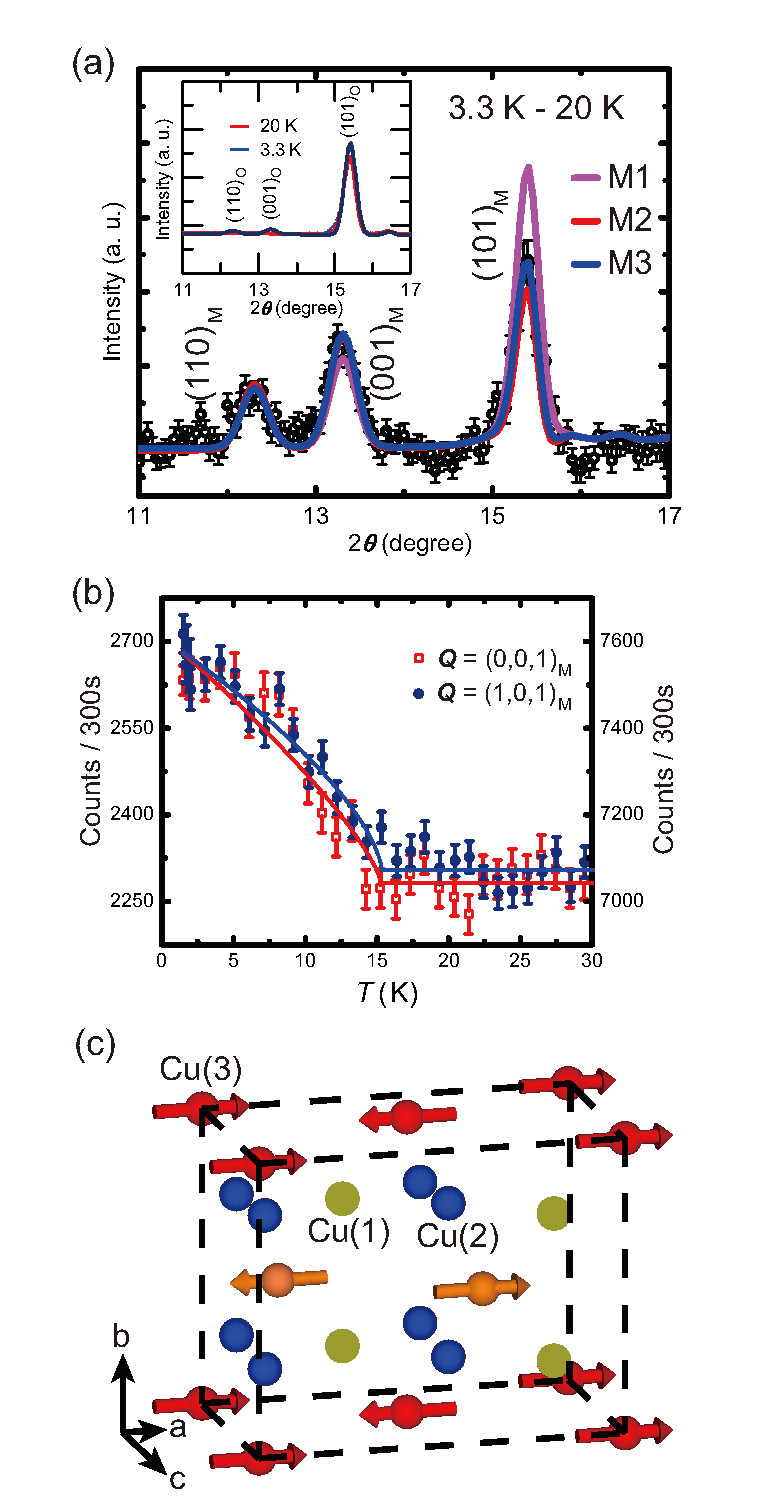}
\caption{(a) The three magnetic peaks obtained by subtracting the 20-K data from the 3.3-K data. The solid lines are calculated results of three types of magnetic structures. The inset shows the refined results of raw data measured at SPODI. The subscript "M" denotes the magnetic structure. (b) Temperature dependence of (0,0,1)$_M$ (open red squares) and (1,0,1)$_M$ (filled blue circles) magnetic peaks measured at HB-2A. The solid lines are the fitted results as described in the main text. (c) Magnetic structure with the arrows indicating the sizes and directions of magnetic moments on Cu(3) ions (red and orange balls). The yellow and blue balls are Cu(1) and Cu(2) ions, respectively, whose structures cannot be determined as discussed in the main text. The dashed black lines indicate the magnetic unit cell with the origin shifted. }
\end{figure}

Figure 2(b) gives the values of the Cu-O bond lengths and Cu-O-Cu angles at 20 K. According to the Goodenough-Kanamori rule~\cite{GOODENOUGH1958,KANAMORI1959}, the nearest-neighbour superexchange changes from positive (antiferromagnetic) to negative (ferromagnetic) when the Cu-O-Cu angle goes through about 95$^{\circ}$~\cite{MizunoY98}. The Cu-O-Cu angles among Cu(3) and three other Cu$^{2+}$ have values both larger and smaller than 95$^{\circ}$ at 20 K, suggesting very complicated superexchange couplings.  

Figure 2(c) shows the temperature dependence of the intensity of the (1,0,2)$_O$ structural peak in the orthorhombic notation, which appears below about 270 K and slowly increases with decreasing temperature until about 200 K. As discussed above, the structural change from high temperature to low temperature is associated with a particular position chosen by Cu(3) from three equivalent positions. Since the position of Cu(3) is randomly distributed among these three positions at room temperature, it follows that Cu(3) ions should be able to resonate among them above 200 K so that their positions are not random any more at low temperatures. At the current stage due to the lack of thermodynamic evidence, we are unable to distinguish whether this structural change is a phase transition or rather a crossover resulting from increasing distortion with decreasing temperature. 

Assuming that there is no structural transition below 20 K, we find that the neutron-diffraction data at 3.3 K [Fig. 1(c)] can be refined by introducing an AF order. To be more explicit, the inset of Fig. 3(a) shows the neutron-diffraction data and the refinement results for the first three peaks with lowest angles. Two new peaks at (110)$_{O}$ and (001)$_{O}$ emerge at 3.3 K, which are forbidden in the orthorhombic structure. The intensity of the (101)$_O$ peak significantly increases from 20 to 3.3 K. The main panel of Fig. 3(a) gives the subtracted results, which clearly shows three magnetic peaks. Figure 3(b) shows the temperature dependence of the intensities of two peaks, which is consistent with the AF transition at about 15 K reported previously \cite{HanTH14}, suggesting their magnetic origin. Within the statistics, only one magnetic transition is observed. With $T_N$ fixed at 15.5 K determined by the specific heat measurements as shown in the next section, the intensities of the (0,0,1)$_M$ and (1,0,1)$_M$ AF Bragg peaks can be fitted by $A(1-T/T_N)^{2\beta}$ with $\beta$ as 0.39 $\pm$ 0.06 and 0.33 $\pm$ 0.04, respectively. These values of the order parameter critical exponent are consistent with the 3D Heisenberg universality~\cite{Campostrini2002}.

With only three magnetic peaks observed, the magnetic structure cannot be unambiguously determined. Symmetry analysis of the different possible magnetic structures is performed using SARAh\cite{WillsAS00}, which gives the propagation vector $k_{19}$ = (0,0,0). Representational analysis on three copper sites gives eight irreducible representations (IR’s), where IR-7 gives the best refinement results. After trying different models, the following conclusions can be made. First, the magnetic unit cell is the same as the orthorhombic one. Therefore, the positions of the (110)$_O$, (001)$_O$ and (101)$_O$ peaks in the inset of Fig. 3(a) are the same as those of the (110)$_M$, (001)$_M$ and (101)$_M$ peaks in the main panel of Fig. 3(a). Second, the moments on all copper sites are always confined within kagome planes, i.e., the ac plane in the orthorhombic structure. Third, the magnetic structures of the Cu(3) ions are always the same with the moment direction along the orthorhombic a axis, as shown in Fig. 3(c). The magnetic moment is about 0.66(7) $\mu_B$, which changes little in different models. 

Although the magnetic configurations of the Cu(1) and Cu(2) ions cannot be determined, we give three examples of refinement results, as shown by the solid lines in Fig. 3(a). In the M1 configuration, the moments at the Cu(1) and Cu(2) ions are fixed to be the same and their directions rotate simultaneously. The value of the moments is found to be about 0.182(35) $\mu_B$ with the moment direction about 30$^{\circ}$ away from the orthorhombic a axis. In the M2 configuration, the moment at the Cu(2) ions is set to zero. The value of the moment is about 0.39(22) $\mu_B$, but the direction is rather arbitrary. In the M3 configuration, the directions of the Cu(1) and Cu(2) moments are set to be along the orthorhombic a axis, the same as that of the Cu(3) moment. The values of the moments on the Cu(1) and Cu(2) ions are 0.329(76) $\mu_B$ and 0.079(37) $\mu_B$, respectively. Although M3 gives the best fit, it cannot be really distinguished from the other configurations within the error bars. In any case, the average moment of Cu(1) and Cu(2) is much smaller than that of Cu(3), which is why the magnetic structure of Cu(3) ions can be settled.  

\subsection{Evolution of the antiferromagnetic order with Zn substitution}

\begin{figure}[tbp]
\includegraphics[width=\columnwidth]{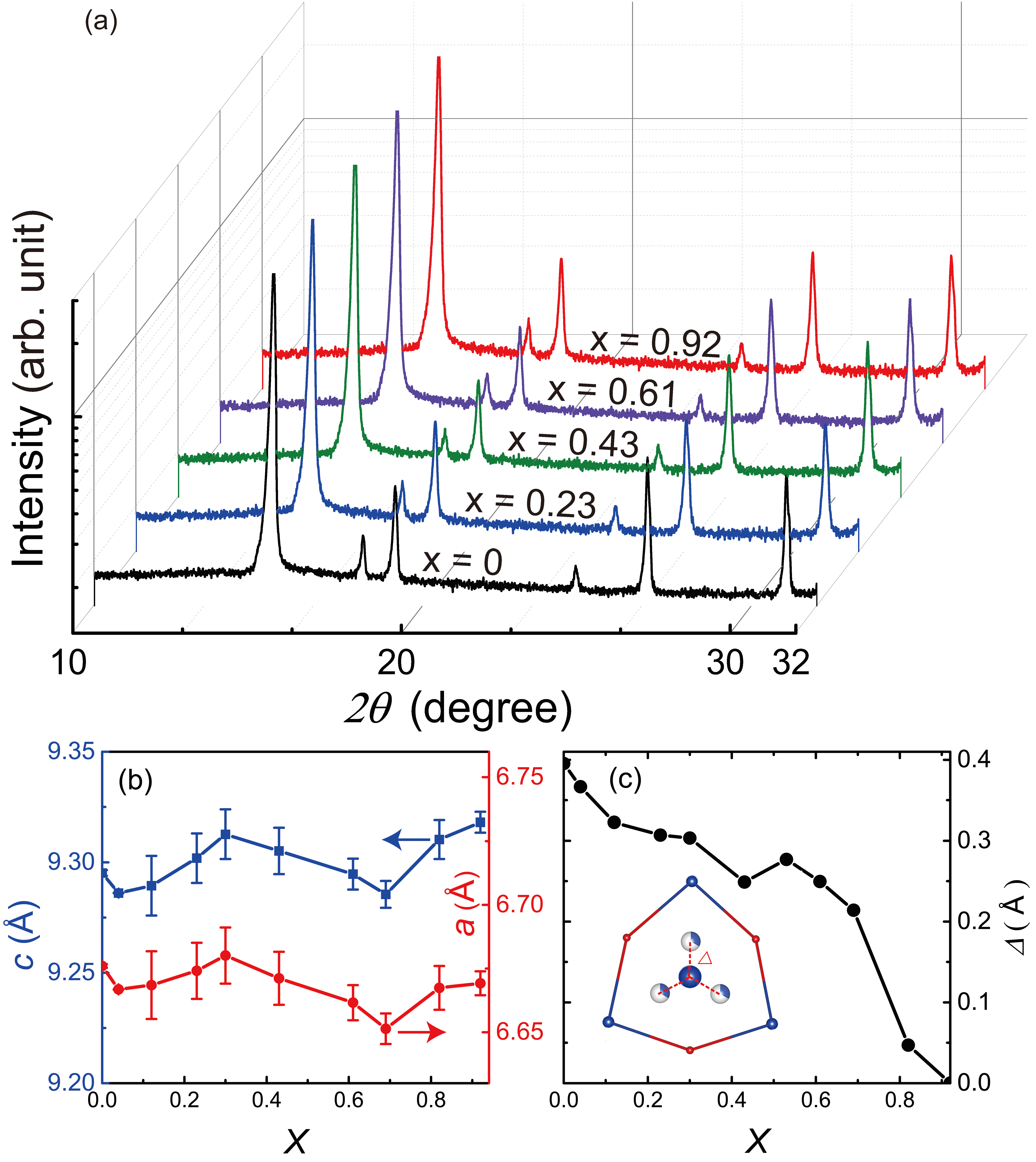}
\caption{(a) Room-temperature x-ray diffraction data of selected Cu$_{4-x}$Zn$_x$(OH)$_6$FBr samples. The intensity axis is plotted with the logarithmic scale. (b) Evolution of lattice constants with Zn substitution level $x$ at room temperature. We note that $a_H$ = b$_H$. (c) Doping dependence of the splitting $\Delta$ of Cu(3) from its average position as defined in the inset.}
\end{figure}

Figure 4(a) shows the room-temperature x-ray diffraction data of Cu$_{4-x}$Zn$_x$(OH)$_6$FBr with different $x$'s from 0 to 0.92. Since there is no new peak appearing with Zn substitution, one can conclude that Zn can be continuously doped into barlowite. All the data can be refined by the hexagonal structure with the space group of $P6_3/mmc$. Figure 4(b) gives the substitution evolution of lattice constants $a_H$ and $c_H$, both of which change little with substitution. As discussed in the previous subsection, the interlayer Cu$^{2+}$ in barlowite has to be refined with three equivalent positions. One can define $\Delta$ as the distance from one of these positions to the center of their average positions as shown by the inset of Fig. 4(c). The value of $\Delta$ continuously decreases with increasing $x$ and becomes zero in the $x$ = 0.92 sample, as shown in Fig. 4(c). The Cu-O-Cu angle between Zn$^{2+}$ or Cu$^{2+}$ at the Cu(3) position and other Cu$^{2+}$ ions on kagome planes of the $x$ = 0.92 sample is 94.73$^{\circ}$ at 4 K according to our previous measurements \cite{WeiY17}, which will give nearly zero superexchange couplings \cite{MizunoY98}. Therefore, it seems that the presence of residual interlayer Cu$^{2+}$ will not affect the spin dynamics of the kagome layers as far as its content is less than 10\%. It should be noted the above analysis has assumed that Zn$^{2+}$ only substitutes interlayer Cu$^{2+}$. At the current stage, we cannot exclude the possibility that a few amounts of Zn$^{2+}$ may substitute Cu$^{2+}$ within the kagome planes \cite{OlariuA08,FreedmanDE10,FuM15}.

\begin{figure}[tbp]
\includegraphics[width=\columnwidth]{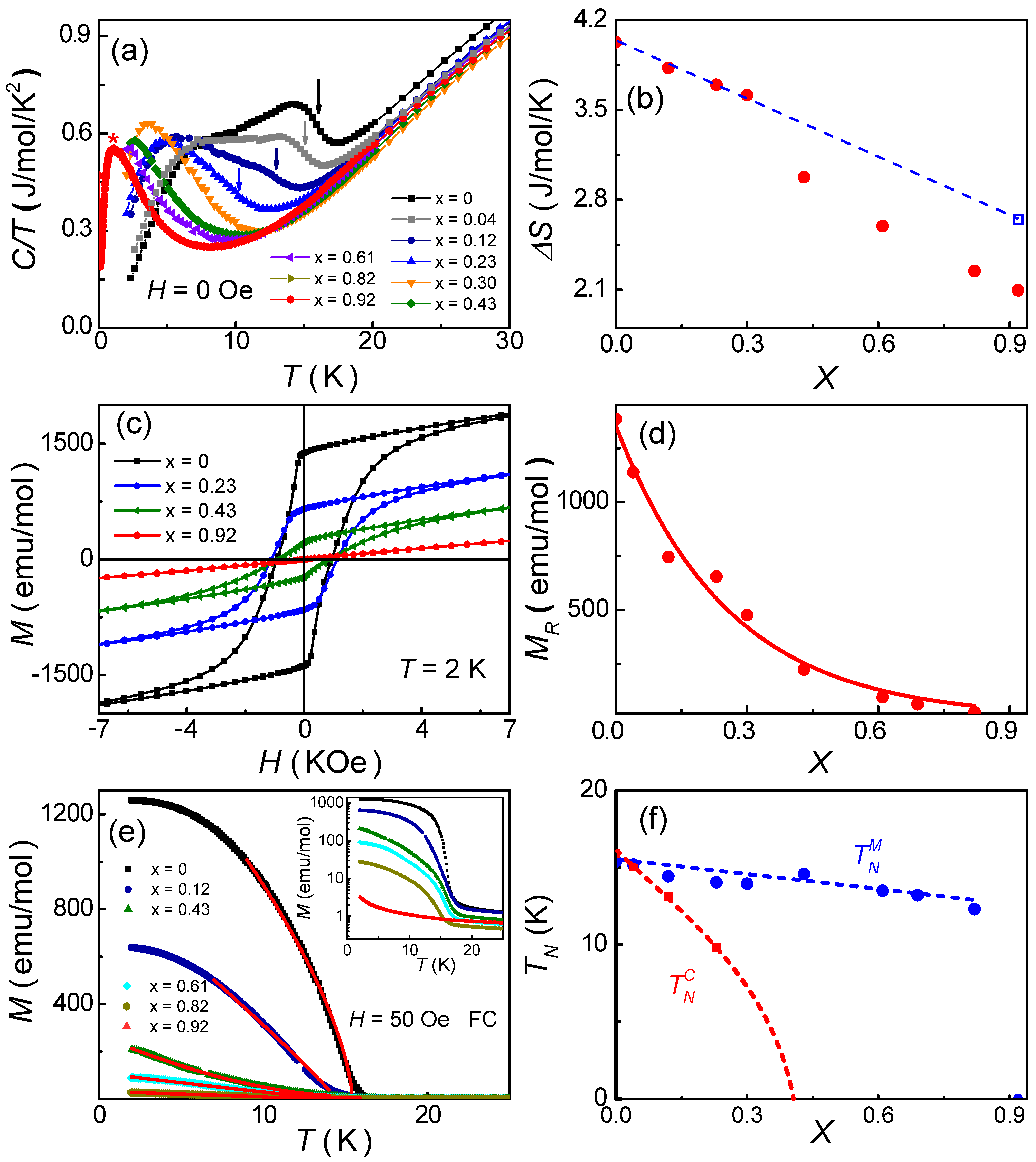}
\caption{(a) Low-temperature specific heat of Cu$_{4-x}$Zn(OH)$_6$FBr. All the samples were measured down to 2 K except for the x = 0.92 sample, which has already been reported previously \cite{FengZL17}. The arrows indicate the bulk magnetic transition temperature $T_N$, which is determined as the middle point of the drop of $\Delta C/T$ during the transition. The low-temperature hump for the x = 0.92 sample is indicated by the star symbol. (b) Temperature dependence of $\Delta S$ as defined in the main text. The red filled circles are results considering $C/T$ above 2 K, whereas the temperature in obtaining the data of the blue open squares is down to 50 mK. (c) Field dependence of magnetization at 2 K for selected samples at low fields. (d) Doping dependence of retentivity $M_R$. The solid line is an exponential fit. (e) Low-temperature magnetic moments of Cu$_{4-x}$Zn$_x$(OH)$_6$FBr measured by the field-cooling process at 50 Oe. The solid lines are fitted results as described in the main text. The inset shows the data with the logarithmic scale. (f) Doping dependence of $T_N$ from specific heat ($T_N^C$, red squares) and susceptibility ($T_N^M$, blue circles) measurements. The $T_N$ of the $x$ = 0.92 sample is manually set as zero. The dashed lines are guides to the eye. }
\end{figure}

The AF transition in barlowite can be observed in specific heat as shown in Fig. 5(a), which is similar as reported previously \cite{HanTH14}. In the x = 0.92 sample, a hump is found in $C/T$ as indicated by the star symbol in Fig. 5(a)\cite{FengZL17} , which has been attributed to the contribution from residual interlayer Cu$^{2+}$ in herbertsmithite\cite{HanTH16}. With decreasing Zn content, the hump does not disappear but moves to higher temperature with little change in the integrated area. On the other hand, starting from the x = 0 sample, the AF transition can be still seen in the $x$ = 0.12 sample, but becomes indistinguishable due to the presence of the hump at low temperatures for $x \geq 0.3$. 

We may estimate the contribution of this hump by analyzing the entropy change $\Delta S$ during the magnetic transition. The high-temperature data are fitted by a simple polynomial function $C_{bg} = \alpha T^2 + \beta T^3$ as performed previously \cite{HanTH14}. The range has been chosen from 30 K to about 5 K higher than $T_N$ for $x<$ 0.23 or 15 K for others. The magnetic part of the specific-heat $C_M$ associated with the AF order can be obtained by subtracting $C_{bg}$ from the raw data. It should be noted that a substantial magnetic contribution to the specific heat should be present above $T_N$ as discussed in the following subsection and the quadratic term in $C_{bg}$ may be associated with it \cite{HanTH14}. $\Delta S$ can thus be obtained by integrating $C_M/T$ from 0 to 20 K, assuming that $C_M/T (T = 0 K) = 0$, which is the case for either the AF ordered or the QSL ground state. However, we have only measured the specific heat down to 2 K for most of the samples as shown in Fig. 5(a). The above process will significantly underestimate the contribution below 2 K for samples with large $x$ since the hump temperature becomes lower than 2 K. We have measured the specific heat of the x = 0.92 sample down to 50 mK, which can give us a more precise value of $\Delta S$ as shown by the blue square in Fig. 5(b). A rough linear substitution dependence of $\Delta S$ from the low-substitution samples to the x = 0.92 sample is found, suggesting that the hump contribution of $\Delta S$ is rather independent of substitution. Therefore, the actual entropy change in $\Delta S$ in barlowite is just about 1.38 J/mol/K, which corresponds to 0.06 $k_B \ln2$ per Cu$^{2+}$, only about one-third of that reported previously \cite{HanTH14}.

Figure 5(c) shows low-temperature field dependence of magnetization $M$ at 2 K. Ferromagnetic-like hysteresis can be found in all the samples except for the $x$ = 0.92 one. The substitution dependence of retentivity $M_R$, i.e., the magnetization at the zero field after the magnetic field is removed, is shown in Fig. 5(d), where $M_R$ decreases exponentially with increasing $x$. The presence of ferromagnetic like hysteresis most likely comes from the domains formed at low temperature due to the orthorhombic structure. The exponential decrease in $M_R$ suggests that either the energy required for overcoming the domain walls or the ferromagnetic-like component of the bulk order decreases quickly with increasing $x$. 

Figure 5(e) shows the temperature dependence of magnetic moment $M$ below 25 K at 50 Oe. The low-temperature signal decreases dramatically with increasing x, but we can still observe magnetic transitions for samples with $x$ up to 0.82. Fitting the data with $A(1-T/T_N)^\beta$ gives the substitution dependence of $T_N^{M}$ in Fig. 5(f), where $T_N^{M}$ only decreases slightly with increasing $x$ and suddenly become zero at $x$ = 0.92. On the other hand, $T_N^{C}$, obtained from the specific heat measurements in Fig. 5(a), decreases quickly with Zn substitution and may becomes zero around $x = 0.4$. It should be pointed out that this value is very subjective and since we cannot observe $T_N^c$ around x = 0.4. Further studies are thus needed to identify the exact substitution level.

\subsection{High-temperature and high-field properties of Cu$_{4-x}$Zn$_x$(OH)$_6$FBr}

\begin{figure}[tbp]
\includegraphics[width=\columnwidth]{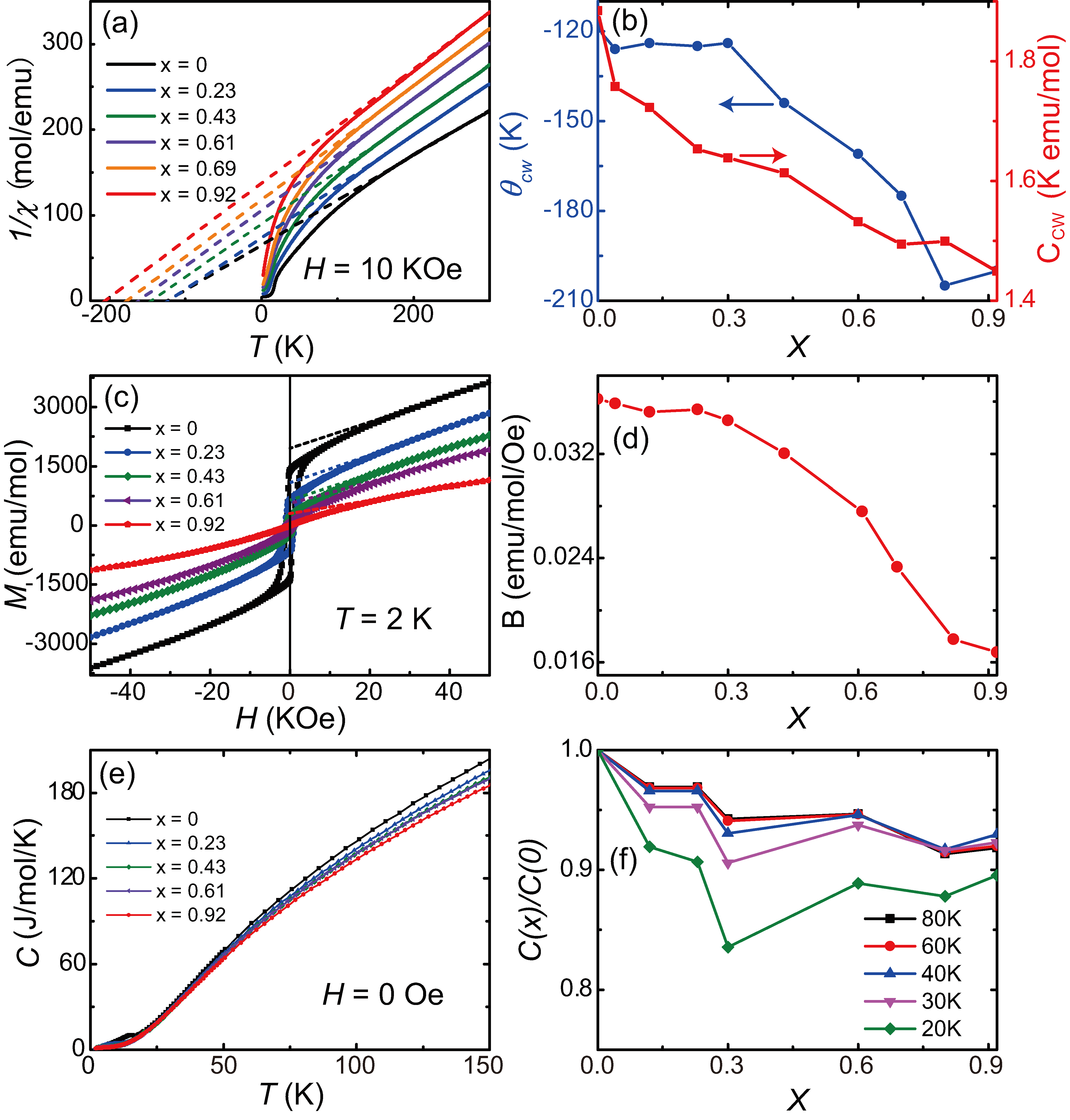}
\caption{(a) Temperature dependence of $1/\chi$ for Cu$_{4-x}$Zn$_x$(OH)$_6$FBr. The dashed lines are fitted results between 150 and 300 K. (b) Doing dependence of the Curie temperature $\theta_{CW}$ (left axis) and Curie constant $C_{CW}$ (right axis). (c) Field dependence of magnetization at 2 K for selected samples at high fields. The dashed lines are linear fitted results from 30 to 50 KOe. (d) Doping dependence of slope B fitted from (c). (e) High-temperature specific heat of selected samples. (f) Doping dependence of $C(x)/C(0)$	 at several temperatures.}
\end{figure}

Figure 6(a) shows the temperature dependence of $1/\chi$ of Cu$_{4-x}$Zn$_x$(OH)$_6$FBr. The high-temperature data from 150 to 300 K can be fitted by the Curie-Weiss function as $\chi = C_{CW}/(T-\theta_{CW})$. The Curie constant $C_{CW}$ decreases linearly with increasing substitution, as shown in Fig. 6(b), which suggests that it is associated with the content of Cu$^{2+}$ ions. However, the absolute value of the Curie temperature $\theta_{CW}$ only starts increasing above $x$ = 0.3, which is consistent with what is observed in the Zn$_x$Cu$_{4-x}$(OH)$_6$Cl$_2$ system \cite{ShoresMP05}. This implies there is a substantial change in the nature of magnetic interactions in the system from $x>0.3$, and we believe it is related to the appearance of the magnetic properties of the kagome QSL plane from $x>0.3$ onwards.

Figure 6(c) shows high-field magnetization at 2 K. The data between 30 and 50 KOe can be fitted by a linear function as $A + BH$. We note that similar measurements on the single crystal of barlowite have shown large anisotropy for fields parallel and perpendicular to the c axis, but the slopes in the above field range are rather the same \cite{HanTH16b}. Figure 6(d) shows substitution dependence of $B$, which starts decreasing above $x=0.3$. This is consistent with the strengthening of antiferromagnetic correlations as indicated by the increasing in $|\theta_{CW}|$. 

Figure 6(e) shows the specific heat data up to 150 K. Apart from low-temperature differences due to the presence of the AF order and hump [Fig. 5(e)], the high-temperature data also show different behaviors. Figure 6(f) plots the substitution dependence of $C(x)/C(0)$, where $C(x)$ and $C(0)$ are the specific heat with Zn substitution levels of $x$ and zero, respectively. This value roughly decreases monotonically with $x$ for $T >$ 60 K, which suggests that the high-temperature specific heat may be dominated by phonons. However, a dip is found at $x$ = 0.3 for those at lower temperatures, indicating that there is a contribution from the kagome QSL plane as suggested in Fig. 6(b) and 6(d). This is also consistent with the observation that only a small amount of entropy is involved during the magnetic transition in barlowite as shown in the previous section.

\section{discussions}

Our results provide a comprehensive picture of the magnetic order in barlowite. The establishment of the AF order is directly associated with the structural distortion at high temperatures, which is the reason that the magnetic structure in Fig. 3 is different from the incorrectly proposed canted antiferromagnetic order based on the first-principles calculation~\cite{JeschkeHO15} without knowing the structural change. It is interesting to note that three different magnetic structures have been proposed for the clinoatacamite Cu$_2$(OH)$_3$Cl \cite{LeeSH07,KimJH08,WillsAS08}. In our case, the magnetic structure in the M3 configuration is consistent with results from bulk measurements on barlowite. For example, the magnetic entropy associated with the magnetic transition is just about 0.06 $k_B \ln2$ per Cu$^{2+}$. In the simplest model, the magnetic entropy is proportional to $M^2$, i.e., the square of the magnetic moment~\cite{Schwablbook}. Here the sum of the ordered moments of four Cu$^{2+}$ ions in the M3 configuration is 1.076 $\mu_B$, whereas the total moment is $4gS$ = 4.54 $\mu_B$, taking $g$ = 2.27 and $S$ = 1/2 \cite{HanTH14}. Therefore, the entropy release above the transition is about  5.6\% of $k_B \ln2$, which is very close to the experimental value. Moreover, in the magnetic measurements on the single crystal, the hysteresis loop is only observed when H$\bot$c but not for H//c as shown in the single crystal measurements \cite{HanTH16b}, which is consistent with our results that the magnetic/structural domains only present within the kagome planes. When the magnetic field is parallel to the c axis, a saturation moment of 0.29 $\mu_B$ per Cu is found \cite{HanTH16b}. Although it is attributed to full polarization of the interlayer Cu$^{2+}$, we find that this value is close to the sum of the ordered moments in the M3 configuration. 

The presence of interlayer Cu$^{2+}$ ions may result in local lattice distortions that will give rise to the magnetic order even when the Zn substitution level is high. This picture is consistent with our observation of magnetic order in magnetic measurements and low-field hysteresis up to $x$ = 0.82, almost substitution independent of $T_N^M$. It also coincides with the nuclear structure refinement results at room temperature, which suggests that the splitting of $\Delta$ of Cu(3) is not zero even when $x$ is as large as 0.82. However, the results from the specific-heat measurements provide another picture where the magnetic order may have already become zero for $x$ larger than 0.4. Since specific heat is a bulk property, it suggests that the magnetic order established at $T_N^M$ is just a secondary phase. This is consistent with our high-temperature and high-magnetic-field results, which suggest that the spin dynamics of the QSL kagome plane may indeed start to appear with Zn substitution level $x >$ 0.3. 

Based on these analyses, the magnetic properties in the Cu$_{4-x}$Zn$_x$(OH)$_6$FBr system can be divided into two parts, the one associated with the kagome planes (and thus bulk) and the one associated with interlayer Cu$^{2+}$ moments (and thus local). In very low substitution samples, the two parts are strongly coupled and cannot be separated, hence the 3D antiferromagnetic order is formed. With increasing Zn substitution, the bulk magnetic order is quickly suppressed and may disappear around $x \sim$ 0.4, but the local magnetic order persists up to $x$ = 0.82 without much change in its $T_N$. It is worth noting that in both the herbertsmithite and the x = 0.92 samples, the spin excitations can be indeed separated into these two independent parts \cite{NilsenGJ13,HanTH16,WeiY17}.

The suppression of bulk magnetic order gives rise to two possible scenarios. In the first one, a magnetic QCP is present around $x \sim$ 0.4, which suggests that Cu$_{4-x}$Zn$_x$(OH)$_6$FBr may provide us the long-thought opportunity to study the quantum phase transition from a magnetic ordered state to a QSL state. On the other hand, it is also possible that the $Z_2$ QSL state in Cu$_3$Zn$_x$(OH)$_6$FBr is very robust against interlayer magnetic impurities so that it may persists up to very low Zn substitution. In this case, the disappearance of the bulk magnetic order may be associated with a first-order quantum phase transition or even phase separation between the 3D AF order and the QSL. It should be noted that since the x = 0 and x =0.92 samples have different structures at low temperatures, it will be interesting to see whether the quantum transition around x = 0.4 happens within the same nuclear structure or not. In any case, the physics is rich and interesting, and further studies are needed to clarify the situation.

\section{conclusions}

Our systematical investigation on the Zn substitution effect on the antiferromagnetism in the kagome Cu$_{4-x}$Zn$_x$(OH)$_6$FBr system have revealed three major conclusions. First, the magnetic order in barlowite is associated with a hexagonal-to-orthorhombic structural change. Second, Zn substitution leads to local lattice distortion and may give rise to phase separation and result in a bulk magnetic order and a local magnetic order. Third, an evolution of spin dynamics on the kagome QSL planes may result in a quantum phase transition around $x=0.4$ between 3D AF order and QSL. Our results suggest that Cu$_{4-x}$Zn$_x$(OH)$_6$FBr is an interesting system and an experimental playground to investigate the intriguing physics of kagome antiferromagnets, and possibly realize the long-thought situation where quantum phase transition between symmetry-breaking and topologically ordered phases. Further works are definitely needed to explore the rich physics in these kagome compounds.

\begin{acknowledgments}
This work was supported by the National Key R\&D Program of China (Grants No. 2016YFA0300502 and No. 2017YFA0302900,2016YFA0300604), the National Natural Science Foundation of China (Grants No. 11874401 and No. 11674406, No. 11374346, No. 11774399, No. 11474330, No. 11421092, No. 11574359 and No. 11674370), the Strategic Priority Research Program(B) of the Chinese Academy of Sciences (Grants No. XDB25000000 and No. XDB07020000, No. XDB28000000), China Academy of Engineering Physics (Grant No. 2015AB03) and the National Thousand-Young Talents Program of China. Research conducted at ORNL’s High Flux Isotope Reactor was sponsored by the Scientific User Facilities Division, Office of Basic Energy Sciences, US Department of Energy. 

Z.F. and Y.W. contributed equally to this work.
\end{acknowledgments}

\end{document}